\documentclass[aps,showpacs,preprint]{revtex4}
\usepackage{epsfig,amssymb,bm,graphics,color,amsmath}
\usepackage{graphicx}
\usepackage{slashed}
\begin{document}

\title{Bulk viscosity of the gluon plasma in a \\
 holographic approach}
\author{R.~Yaresko, B.~K\"ampfer}
\address{Helmholtz-Zentrum Dresden-Rossendorf, Institute for Radiation Physics,
P.O.~Box 510119, 01314 Dresden, Germany \\
and\\
TU Dresden, Institut f\"ur Theoretische Physik, 01062 Dresden, Germany}

\pacs{11.25.Tq, 47.17.+e, 05.70.Ce, 12.38.Mh, 21.65.Mn}

\begin{abstract}
A gravity-scalar model in 5-dim.\ Riemann space is adjusted to the thermodynamics 
of SU(3) gauge field theory in the temperature range 1 - 10 $T/T_c$ to
calculate holographically the bulk viscosity in 4-dim.\ Minkowski space.
Various settings are compared,
and it is argued that, upon an adjustment of the scalar potential
to reproduce exactly the lattice data within a restricted temperature interval
above $T_c$, rather robust values of the bulk viscosity to entropy density 
ratio are obtained.
\end{abstract}

\maketitle

\section{Introduction}

The duality of ${\cal N} = 4$ supersymmetric Yang-Mills theory in 4-dim.\
Minkowki space with 
type IIB superstring theory on $AdS_5 \times S^5$  
\cite{AdS_CFT} has initiated a wealth of investigations aimed at exploiting the
AdS/CFT correspondence to relate mutually properties of the gravitation sector
(which is anti-de Sitter (AdS) in 5-dim.\ Riemann space) 
with conformal field theories (CFT). Such techniques look particularly useful for
4-dim.\ strongly coupled theories, where real-time processes are difficult to
access. This in turn applies especially to strongly interacting systems, as subjects to QCD,
created in the course of relativistic heavy-ion collisions, i.e.\ the quark-gluon plasma (QGP).      
Here, holographic techniques, based on the AdS/CFT correspondence,
allow to calculate from suitable gravity duals the wanted observables quantifying
properties of the QGP.
Among the important quantities is the bulk viscosity which has a potentially
strong impact on the analysis of the flow pattern in relativistic heavy-ion collisions
\cite{HIC_flow} and may help to solve the photon-$v_2$ puzzle \cite{photon_v2}.

The gravity dual of QCD, even in the pure Yang-Mills sector, is not known. Moreover,
QCD is not a CFT since, due to dimensional transmutation, an inherent energy scale is
emergent which steers the running coupling. In such a situation and 
with a lacking top-down approach from string theory, it looks promising to utilize
a bottom-up approach which incorporates a selected set of properties one is going
to calculate after an appropriate adjustment of the 5-dim.\ Einstein gravity theory
which emerges, strictly speaking, only 
in the large-$N_c$ limit and at large 't Hooft coupling. 
A famous example is the gravity-scalar set-up, where a real scalar field is
consistently coupled to gravity. 
The scalar $\phi$, dual to an 
operator $\cal O_{\phi}$, breaks conformal invariance of AdS space, 
simulating the corresponding breaking 
in Yang-Mills theory, the latter being 
expressed by the trace anomaly relation 
$T_\mu^\mu = \beta(\alpha)/(8\pi \alpha^2) \text{Tr}\, F^2$ 
of the Yang-Mills energy-momentum tensor $T_{\mu\nu}$, $\beta$ function, 
running coupling $\alpha$ and trace of the field strength tensor squared $\text{Tr}\, F^2$. 
Being interested in thermodynamic
properties of the gluon plasma one embeds in the asymptotically AdS space a black brane
which introduces a temperature via Hawking temperature and an entropy
via Bekenstein-Hawking entropy. Besides the equilibrium thermodynamics, encoded
in the gravity metric as dual of the gauge theory energy-momentum tensor, near-to-equilibrium
quantities are accessible as correlators based on the energy-momentum tensor.
For a medium without conserved charges these are the shear and bulk viscosities as 
first-order transport coefficients in a gradient expansion. 

\section{Gravity-scalar holographic models}

The class of gravity-scalar duals is defined by the action 
\begin{equation}
S = \frac{1}{16 \pi G_5} \int d^5 x \sqrt{-g}
\left( R - \frac12 (\partial \phi)^2 - V(\phi)\right) + {\cal L}_{GH} 
\end{equation}
where ${\cal L}_{GH}$ is the Gibbons-Hawking surface term,
irrelevant or our purposes, and $G_5$ denotes the 5-dim.\ gravity constant.
The "potential" $V(\phi)$ determines the self-interaction of the scalar $\phi$; it
contains the constant term $V_0 = -12 / L^2$ ensuring asymptotic AdS behavior with $L$ 
being the curvature scale set by the negative cosmological constant.   
The Riemann space is accordingly specified by extending the conformally
flat 4-dim.\ space-time by the bulk variable $u$ resulting in the ansatz for the
infinitesimal line element squared
\begin{equation}
ds^2 = \exp\{2 A(u)\} \left (d\vec x^2 - f(u) dt^2 + \frac{1}{f(u)} du^2 \right),
\end{equation}  
where (in conformal coordinates) $\lim_{u \to 0} f(u) = 1$ and
$ \lim_{u \to 0} A = \log (L / u)$ ensure the AdS property at the boundary 
$u \to 0$ and
the simple zero of $f(u_H)$ defines the horizon at $u_H > 0$. 

The scalar is supposed to have a radial profile $\phi (u)$ which for 
potentials such as 
$V(\phi) = V_0 + \frac12 m^2 \phi^2 + \cdots$ 
is constrained by the equation of motion to 
$\phi(u) = \phi_{(4 -\Delta)} u^{4 -\Delta} + \phi_\Delta u^\Delta + \cdots$
near the boundary of AdS, where $\phi_{(4 -\Delta)}$ implies an additional term
$\propto \int d^4 x \phi_{(4 - \Delta)} {\cal O}_\phi$ as deformation of the 
original CFT and $\langle {\cal O}_\phi \rangle \propto \phi_\Delta$,
i.e.\ $\phi$ is holographically dual to the operator ${\cal O}_\phi$ with
conformal dimension $\Delta_\phi$. For $\Delta_\phi = 4$, the dual operator 
is exactly marginal and the scalar field is massless, while for $\Delta_\phi \ne 4$
the source $\phi_{(4 - \Delta)}$ introduces a mass scale
$\Lambda = \phi_{(4 - \Delta)}^{1/(4-\Delta)}$ which explicitly breaks
conformal invariance. 
The mass $m$ and the conformal dimension $\Delta_\phi$ are related by $m^2 L^2 = \Delta_\phi (\Delta_\phi -4)$, 
which must satisfy $m^2 L^2 \ge -4$ to fulfill the Breitenlohner-Freedman bound. Renormalizability on the gauge theory 
side requires $\Delta_\phi \leq 4$, i.e.\ $m^2L^2 \leq 0$. While an extension to $1 \leq \Delta_\phi \leq 2$ is possible 
\cite{AdSCFT_symmbreak}, we restrict our attention to the upper branch of the mass-dimension relation and relevant operators, 
i.e.\ $2 < \Delta < 4$. 
This is already a special setting
which follows, e.g., \cite{Gubser_review,Gubser,Gubser_PRL} and serves as outline of our analysis below. 
The improved holographic QCD (IHQCD) model \cite{Kiritsis},
in contrast, is based on different potential asymptotics 
$V(\phi) - V_0 \propto e^\phi + \cdots$ which encodes the running 't Hooft coupling $\lambda \propto e^\phi$
close to the boundary (here at $\phi \rightarrow -\infty$) and 
results in the marginal case $\Delta_\phi = 4$, while, for large 't Hooft coupling, $V(\phi)$ is 
constructed to accomodate confinement and a linear glueball spectrum, cf.\ \cite{Gubser_review, Nitti}.

\section{Thermodynamics}

The two basic AdS/CFT thermodynamic relations
\begin{equation}
T = \left. - \frac{1}{4 \pi}\frac{d f}{du} \right\vert_{u_H}, 
\quad
s= \frac{1}{4 G_5} \exp\{ 3 A\} \vert_{u_H}
\end{equation}
determine the thermodynamics, e.g.\ by $s(T) /T^3$ for parametrically given
temperature $T(u_H)$ and entropy density $s(u_H)$. 
Here, $u_H$ is the horizon position in the bulk.
Einstein's equations determine, via the above conditions at the boundary,
the metric coefficients at $u_H$.
To be specific we utilize 
\begin{equation} \label{eq:V}
V(\phi) L^2 = -12\cosh \gamma \phi + b \phi^2 +
\sum_{n=2}^{5} c_{2n} \phi^{2n}
\end{equation}
with $b = 6\gamma^2 + \Delta(\Delta-4)/2$ from \cite{Gubser, Gubser_PRL}, but 
use solely the matching
condition to lattice data of the SU(3) Yang-Mills equation of state in a finite temperature interval above $T_c$.
That is we ignore an {\it a priori} scale setting at a certain energy and 
leave thus $2 < \Delta < 4$, $\gamma$ and $c_{2n}$ as free parameters.
 
Without further integration constant, the velocity of sound squared,
$v_s^2 = d \log s / d \log T$, is given, while the pressure 
$p = p_c +\int_{T_c}^{T} dT' \, s(T')$, 
energy density $e = -p +sT$ and interaction measure $I = e- 3p$
need one additional constant. A possibility is to employ the lattice
input with $p_c = p(T_c)$, which needs a definition of $T_c$.
The IHQCD model has a clear definition of $T_c$; other options could
be to choose $T_c = T_{min}$, where 
$T_{min}$ is the minimum 
of the temperature $T$ as a function of $u_H$ or $s/T^3$; in the latter 
case, the inflection point $T_{ip}$ can be utilized to define $T_c$ in cases where 
$T$ as a function of $s/T^3$ does not have a minimum.
If one refrains to catch Yang-Mills features at zero temperature
(e.g.\ a linear glue ball spectrum w.r.t.\ a radial quantum number) 
and the latent heat in the deconfinement phase transition 
as in IHQCD \cite{Kiritsis}
one can adjust the value of $T_c$
arbitrarily; also, $G_5$ can be chosen without other constraints than
the optimum reproduction of a given data set in a restricted temperature
interval above $T_c$.     
Here, we choose $LT_c = (LT_{min}, LT_{ip})$ and adjust $\gamma$, $\Delta$, $c_{2n}$ and $G_5/L^3$ by minimizing
\begin{equation} \label{eq:chi}
\chi^2_{s/T^3} = 
\log \left(\frac{1}{N} \sum_{i=1}^N  \Big[ \sigma(x_i) - y(x_i T_c L) \Big]^2 \right),
\end{equation}
where $\sigma \equiv s(T) / T^3$ refers to the lattice data at $N$ mesh
points $x_i \equiv T_i /T_c$
and $y \equiv G_5 s(T L) /(T L)^3 $ to the holographically calculated
scaled entropy density.

\section{Bulk viscosity}

The class of gravity-scalar models considered here belongs to so-called
two-derivative models which provide the normalized shear viscosity $\eta / s  =1/(4\pi)$,
irrespectively of a specific form of $V(\phi)$, at variance with the asymptotic
behavior of weakly coupled QCD \cite{AMY} and  the expected minimum near $T_c$.
Higher-order gravity models \cite{higher_orders}
abandon such a temperature independence.
Nevertheless, in the strongly coupled region, $\eta / s  =1/(4\pi)$ 
represents an intriguingly important result 
which got popular since the analysis of flow observables
in relativistic heavy-ion collisions at RHIC and LHC appeared consistent
with that.

The bulk viscosity $\zeta$ follows within the present set-up from 
\begin{equation}
\frac{\zeta}{\eta} = \Bigg(\frac{d\log V}{d\phi}\Bigg)^2 
\vert p_{11}\vert^2 \Big\vert_{\phi_H}
\end{equation}
where (using the profile of the scalar field as bulk coordinate) the horizon value of the perturbation $p_{11}$ 
of the $x_1 x_1$-metric component is determined by solving a linearized Einstein equation \cite{Gubser_viscosity}.
\subsection{Optimum adjustment to lattice data}

\begin{figure}
\begin{center}
\includegraphics[width=0.49\columnwidth]{{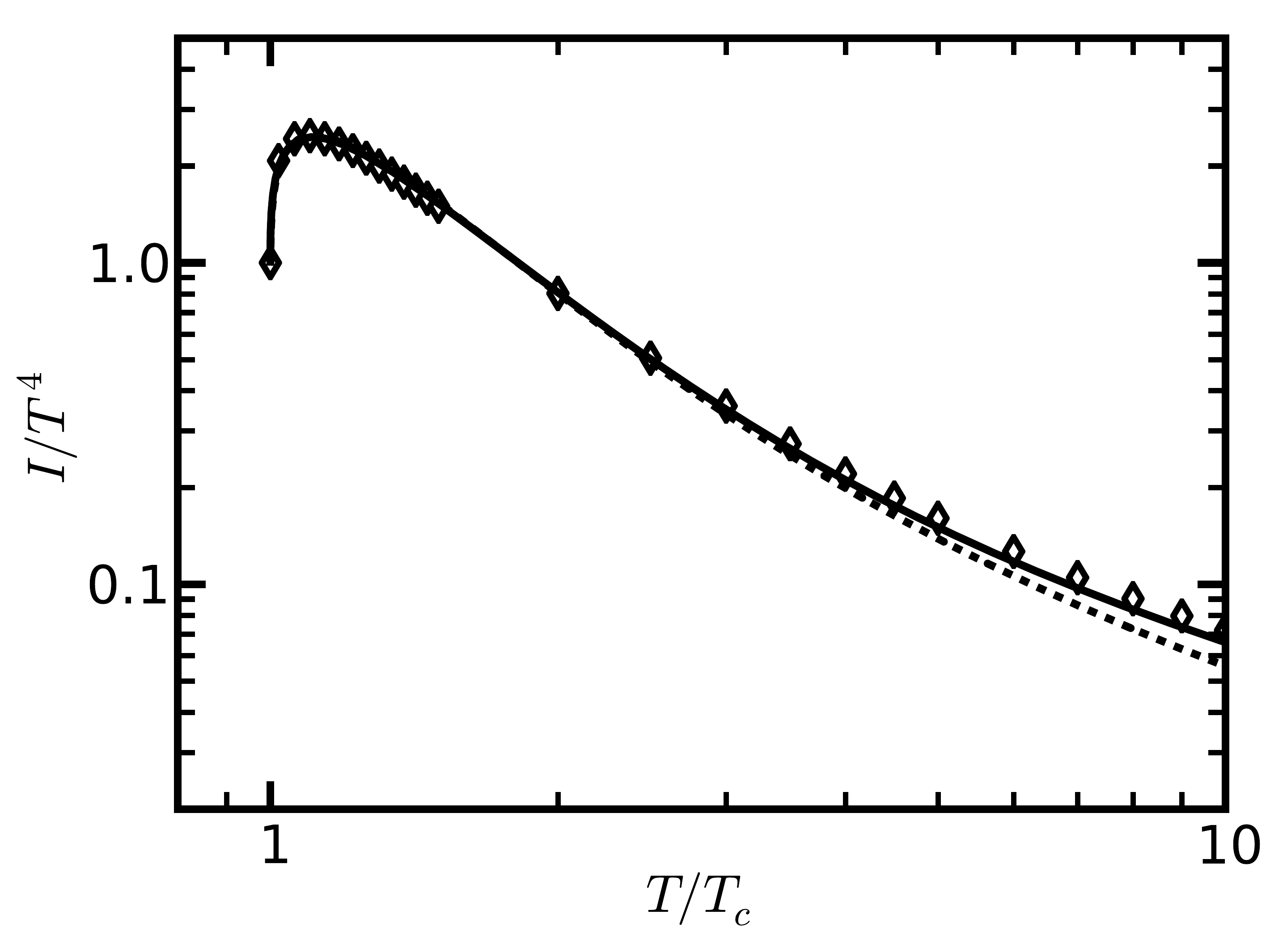}}
\includegraphics[width=0.49\columnwidth]{{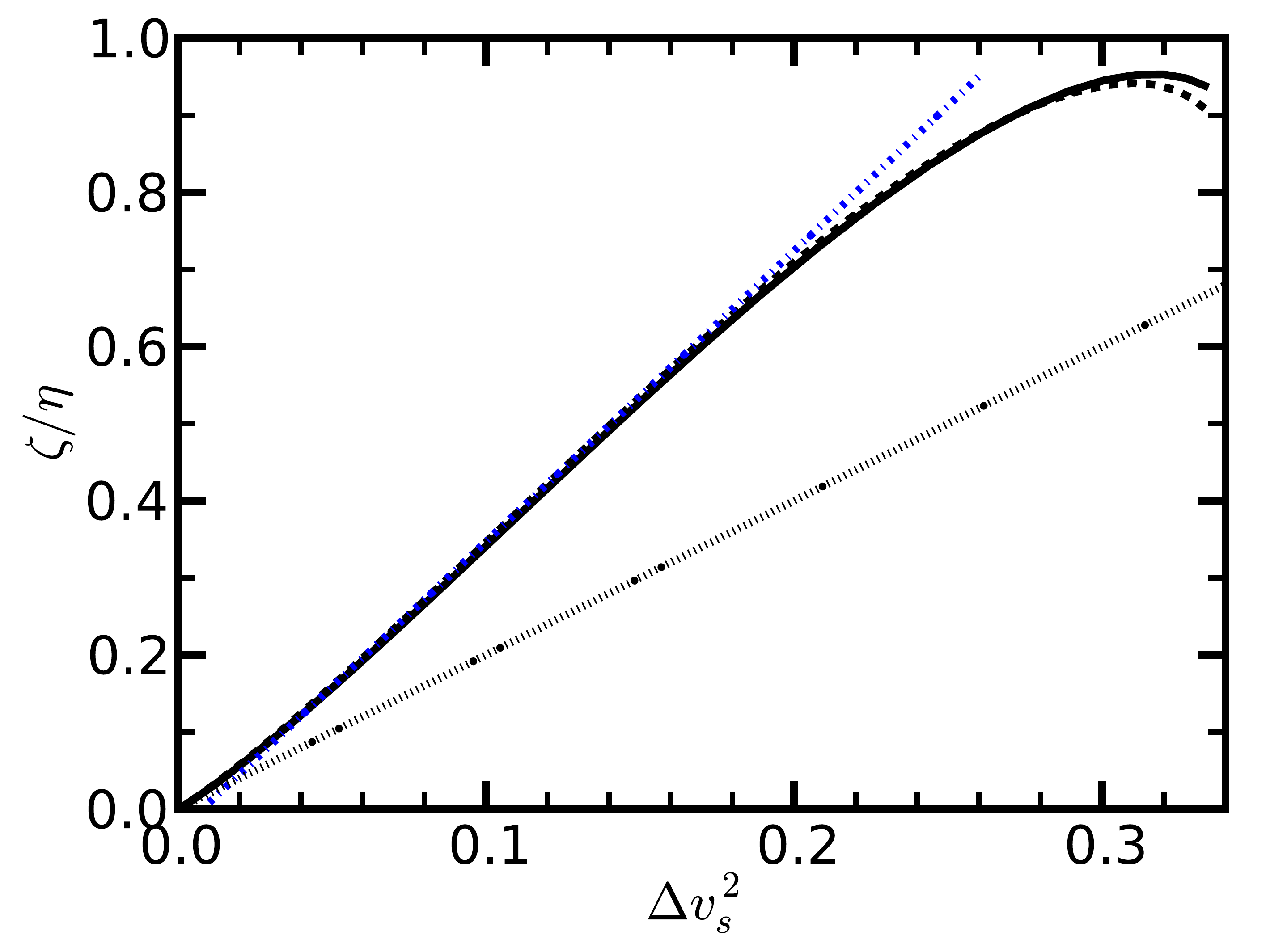}}
\end{center}
\caption{Left: Scaled interaction measure as a function of $T/T_c$. The
solid (dashed) curve is for the potential (\ref{eq:V}) with (without) the
polynomial distortions $c_{2n} \phi^{2n}$. Other thermodynamic quantities
(e.g.\ $v_s^2$, $e/T^4$, $p/T^4$ and $s/T^3$) agree perfectly (cf.\ \cite{RY_BK}) with the
lattice data (symbols, from \cite{Wuppertal_Bp}).
Right: Bulk to shear viscosity ratio as a function of the non-conformality measure. The blue 
dot-dashed line is a linear fit $\zeta/\eta = 1.2\pi \Delta v_s^2 - 0.03$, 
while the dotted line depicts the Buchel bound $\zeta/\eta = 2 \Delta v_s^2$ \cite{Buchel_bound}.
}
\label{fig:1}
\end{figure} 

As shown in \cite{RY_BK}, a perfect matching to lattice data is accomplished 
by the potential (\ref{eq:V}) for $\Delta = 3.7650$ and $\gamma = 0.6580$
when including the polynomial distortions $c_{2n}$; omitting the latter ones 
(with $\Delta = 3.5976$ and $\gamma = 0.6938$) 
the match is near-perfect, see left panel in Fig.~\ref{fig:1}. The bulk to shear viscosity ratio 
(cf.\ right panel in Fig.~\ref{fig:1}) 
displays a linear section, where $\zeta / \eta = \pi C \Delta v_s^2$ with
$C \approx 1.2$, thus fulfilling the Buchel bound $\zeta / \eta \ge 2 \Delta v_s^2$ \cite{Buchel_bound}. 
Such a linear relation $\zeta / \eta \propto \Delta v_s^2 = 1/3 - v_s^2$ 
is considered in \cite{Wiedemann_review} as interesting but as unclear
whether it is a generic result of $Dp$ brane gauge theories. With the results of the
next subsection we argue that it is generic for the gravity-scalar set-up only
for perfect matching to SU(3) Yang-Mills theory. We emphasize that a quasi-particle model 
\cite{Bluhm}
obeys quantitatively a similar proportionality in the strong couping regime, also with
the perfect matching of SU(3) Yang-Mills thermodynamics as a prerequisite.

\subsection{Dependence of bulk viscosity on potential parameters}
 
\begin{figure}
\begin{center}
\includegraphics[width=0.8\columnwidth]{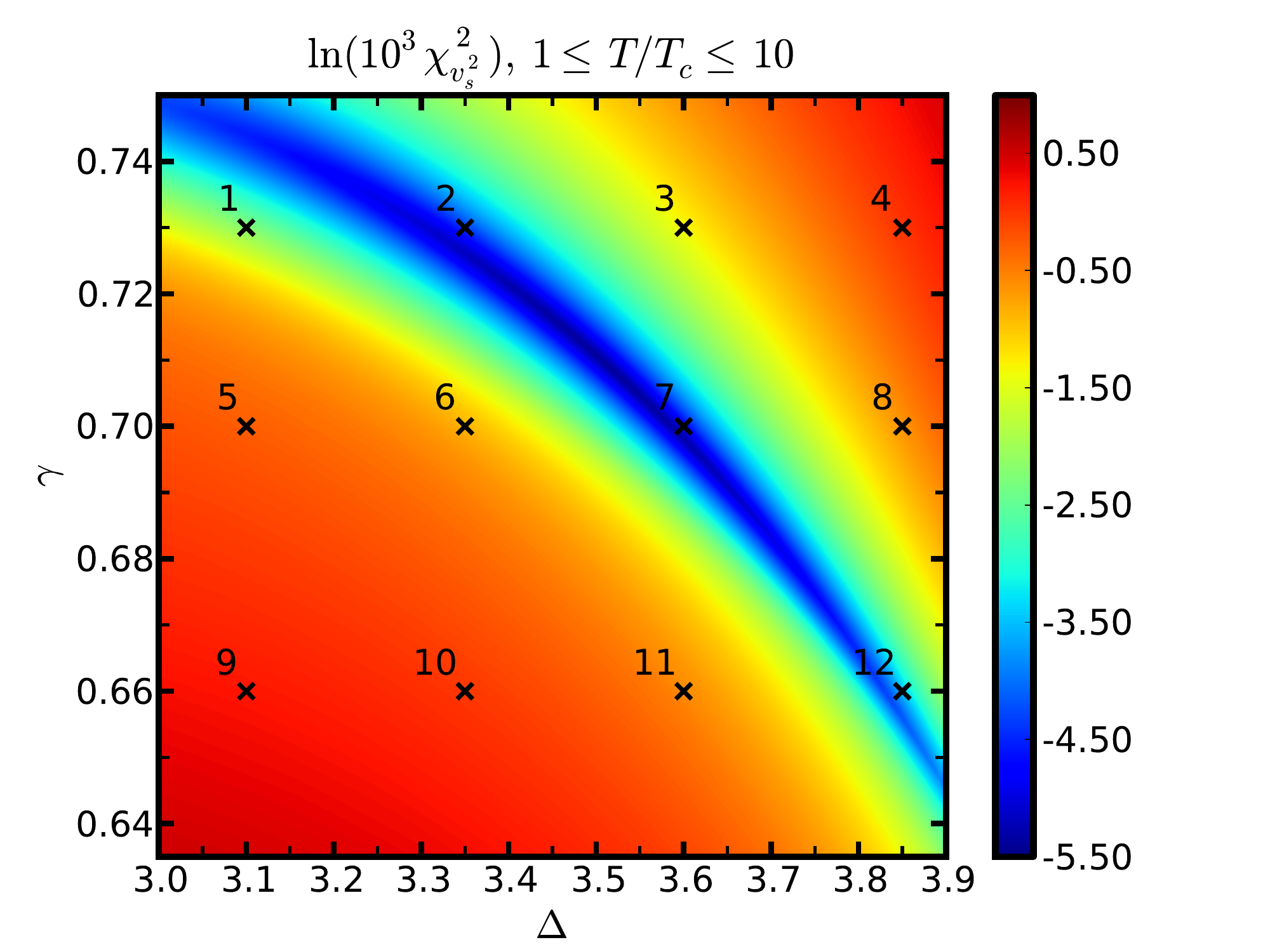}
\end{center}
\caption{The $\chi_{v_s^2}^2$ landscape over the $\gamma$ vs.\ $\Delta $ plane.
The numbers indicate loci of selected parameter choices to be analyzed. 
}
\label{fig:2}
\end{figure}

We demonstrate now the sensitivity of the bulk viscosity on the parameters
of the potential (\ref{eq:V}) with $c_{2n} = 0$. The analysis is restricted to $3 \leq \Delta \leq 3.9$. 
The numbers in Fig.~\ref{fig:2} indicate selected loci at which we
calculate the equation of state and the bulk viscosity
exhibited in Fig.~\ref{fig:3} below. 
The deviation measure $\chi^2_{v_s^2} = \frac{1}{N} \sum_{i=1}^N [v_s^2(x_i) - v_{s,L}^2(x_i T_c L) ]^2$ 
indicates already the (in)accuracy of matching 
the velocity of sound squared, $v_s^2$, from lattice QCD. 
Hereby, $v_s^2$ and $v_{s,L}^2$ are obtained from the holographic 
calculation and the lattice data; $x_i$ and $LT_c$ are as in \eqref{eq:chi}. 
We emphasize the corridor, in which the points 2, 7 and 12 
are localized, which deliver an equally good, though not perfect, reproduction of the lattice data 
(cf.\ left column of Fig.~\ref{fig:3}), due to the individual adjustments of $G_5$. The values of $\zeta/T^3$ spread 
out by a factor of three for $T > T_c$ when comparing the results for all considered loci 1 - 12 
(cf.\ middle column of Fig.~\ref{fig:3}). In contrast, $\zeta/\eta$ as a function of the 
non-conformality measure $\Delta v_s^2$ looks very much the same for loci 2, 7 and 12, 
while for the other loci significant variations of $\zeta/\eta$ can be observed, in particular 
for $\Delta v_s^2 \rightarrow 1/3$, i.e.\ for $T \rightarrow T_c$. This observation lets us 
argue that a perfect matching of the equation of state may lead to a robust result for $\zeta/\eta$.

\begin{figure}
\begin{center}
\includegraphics[width=0.327\columnwidth]{{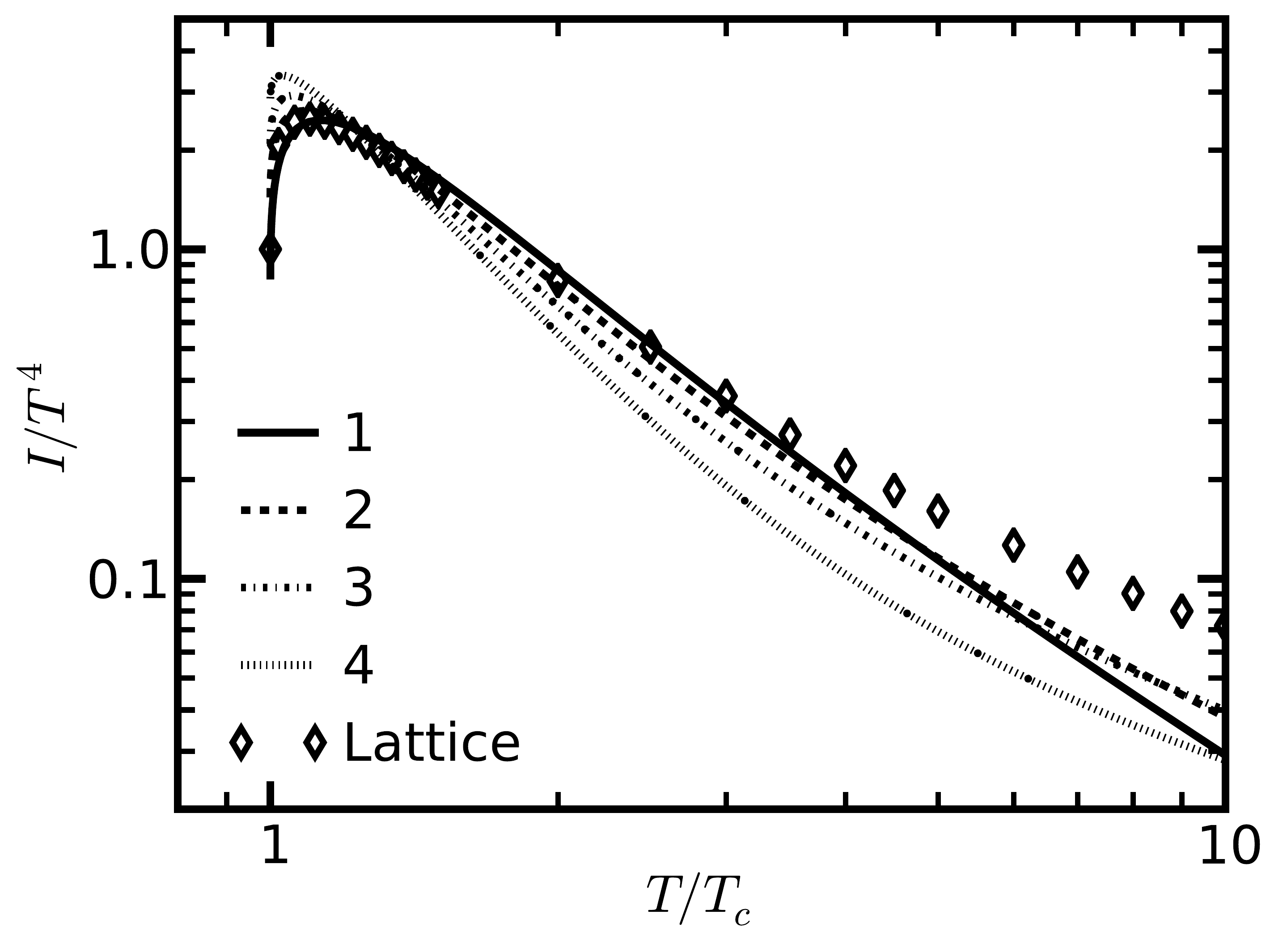}}
\includegraphics[width=0.327\columnwidth]{{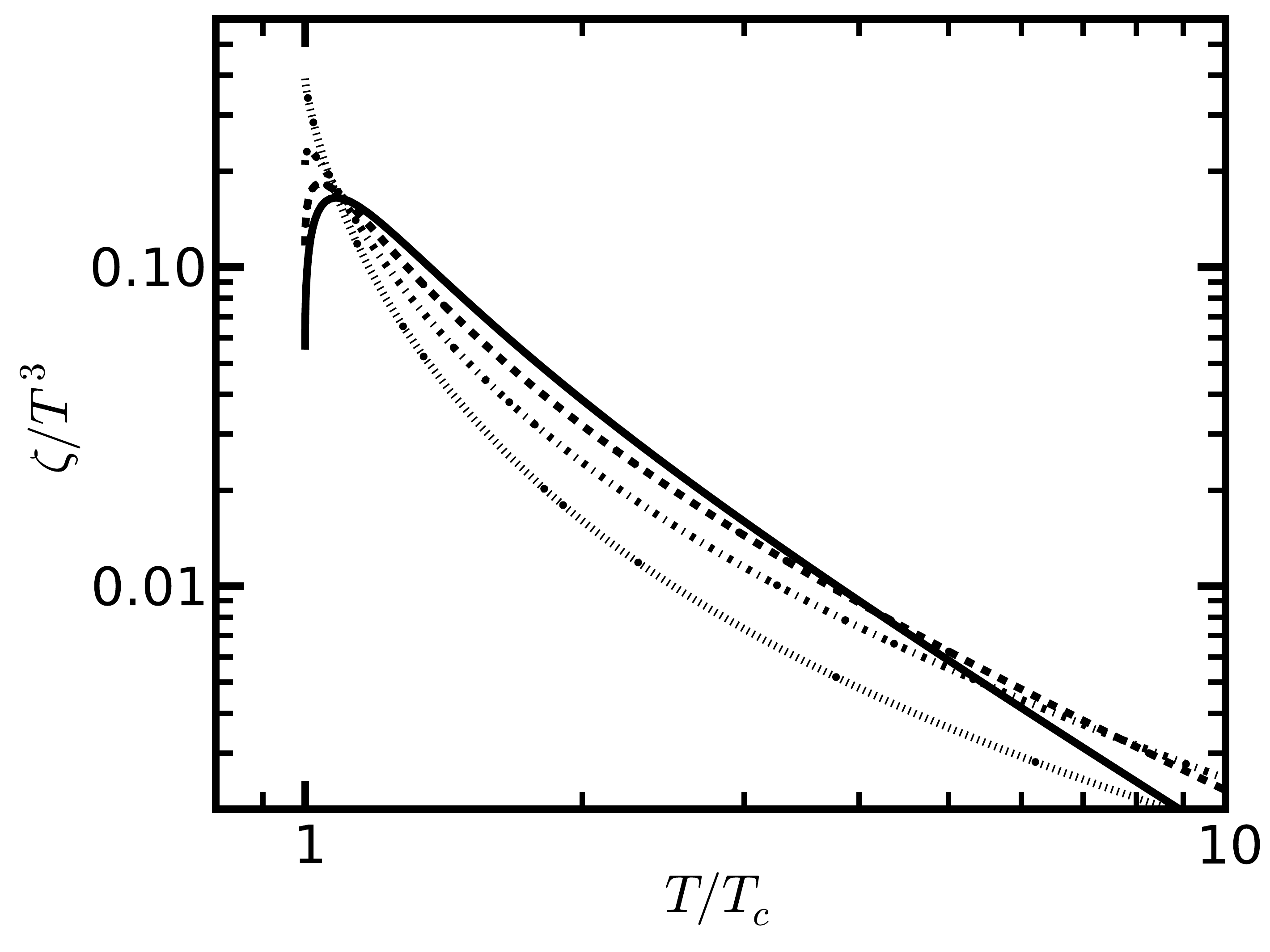}}
\includegraphics[width=0.327\columnwidth]{{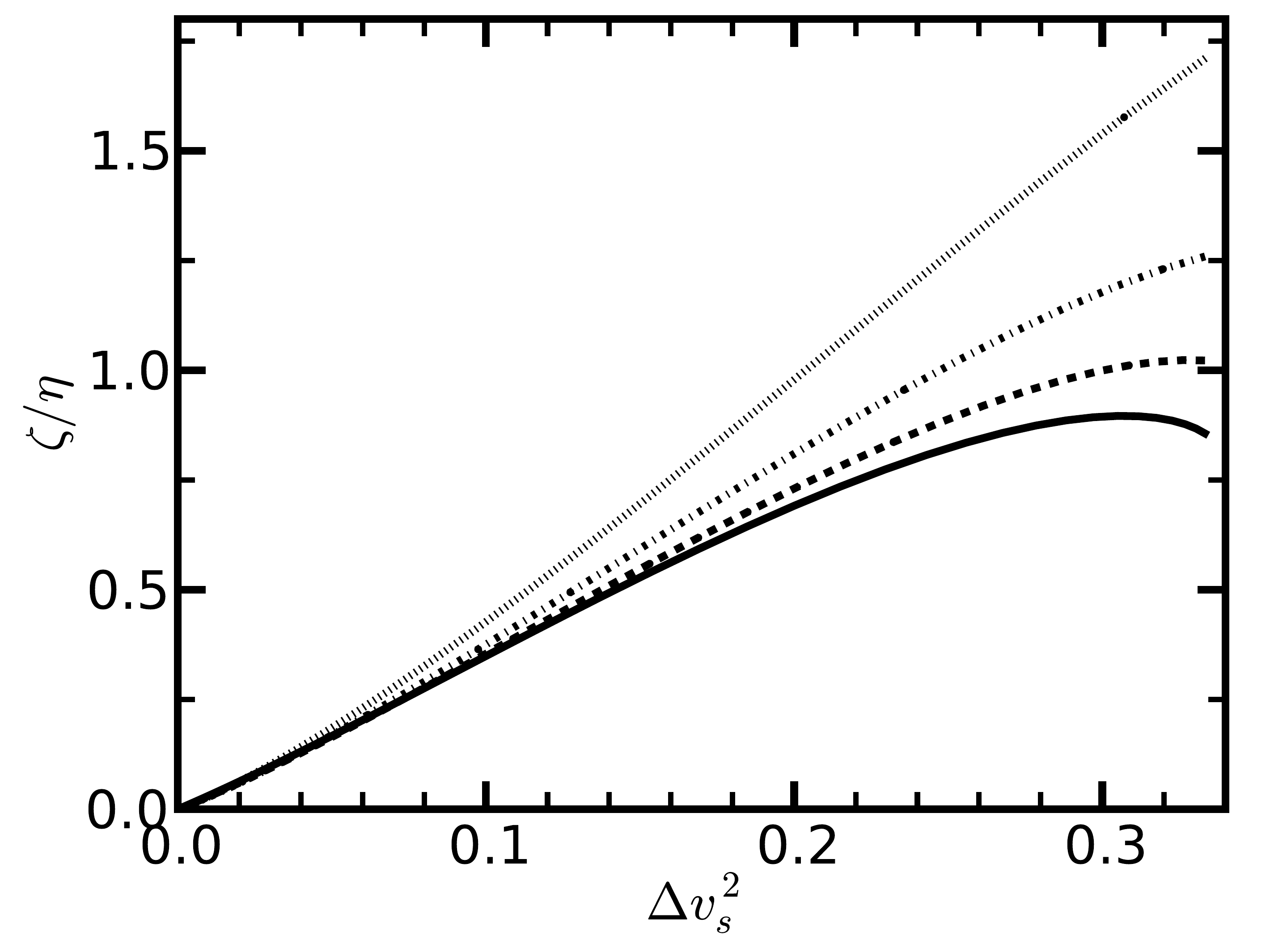}}

\includegraphics[width=0.327\columnwidth]{{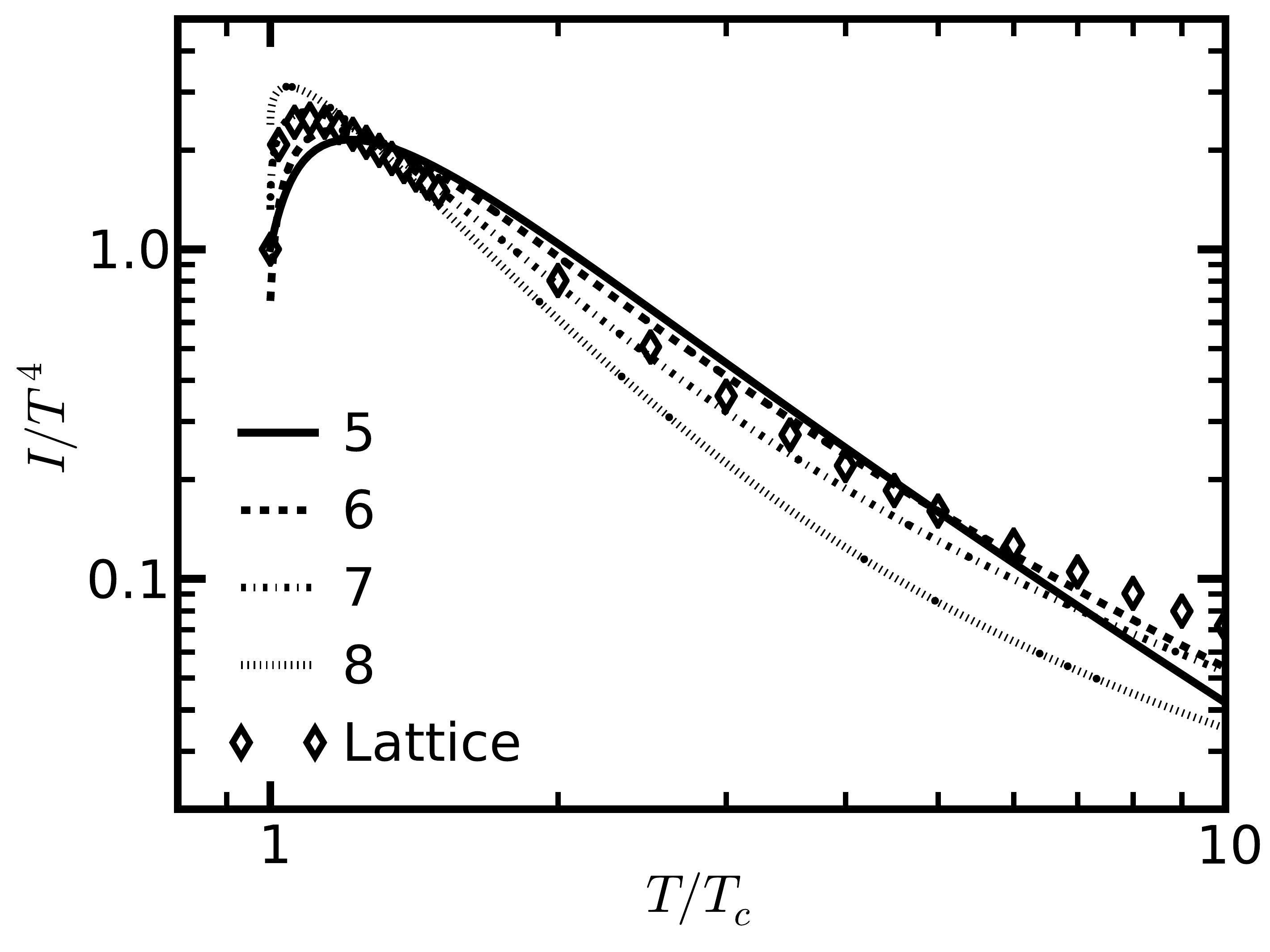}}
\includegraphics[width=0.327\columnwidth]{{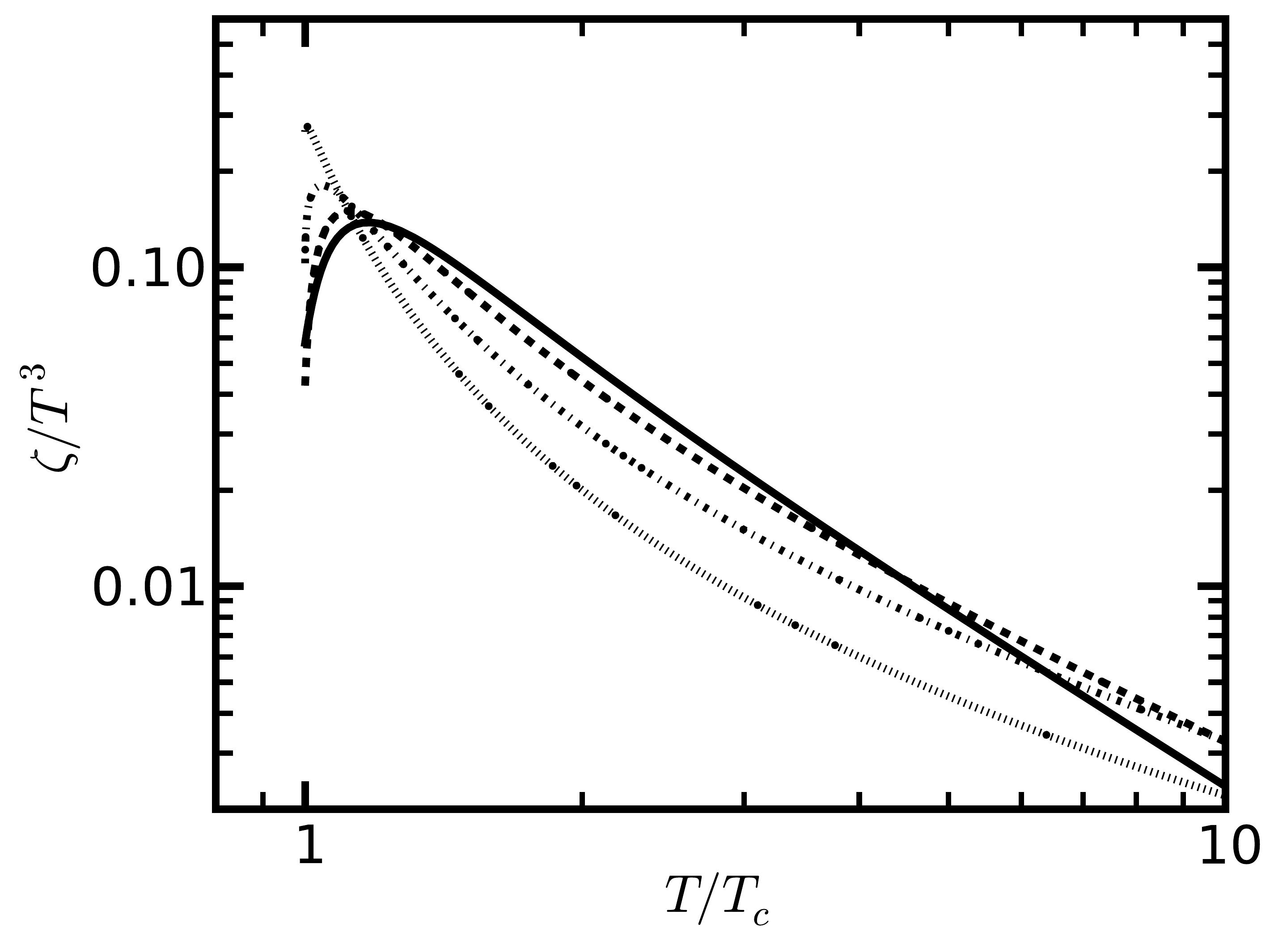}}
\includegraphics[width=0.327\columnwidth]{{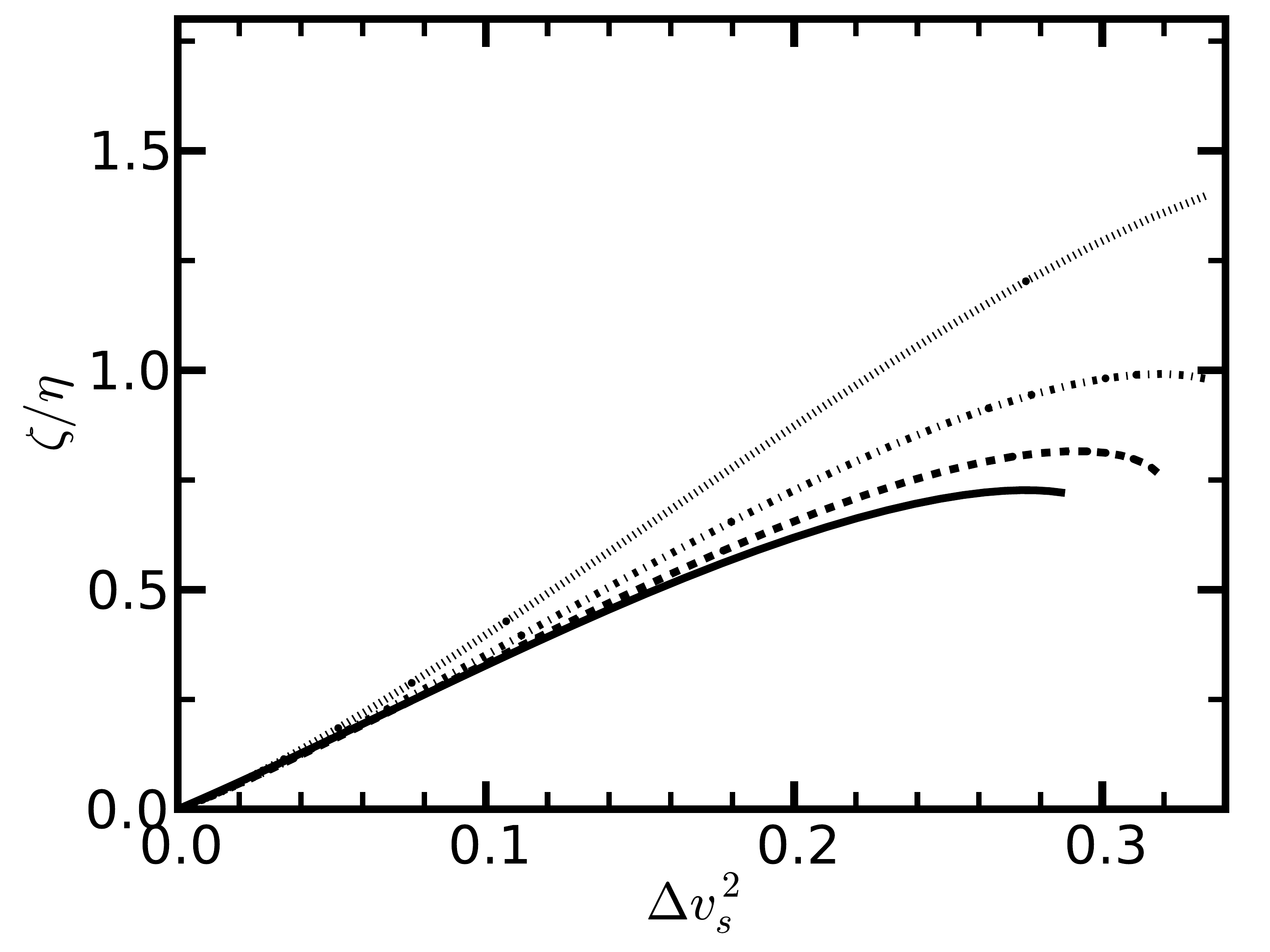}}

\includegraphics[width=0.327\columnwidth]{{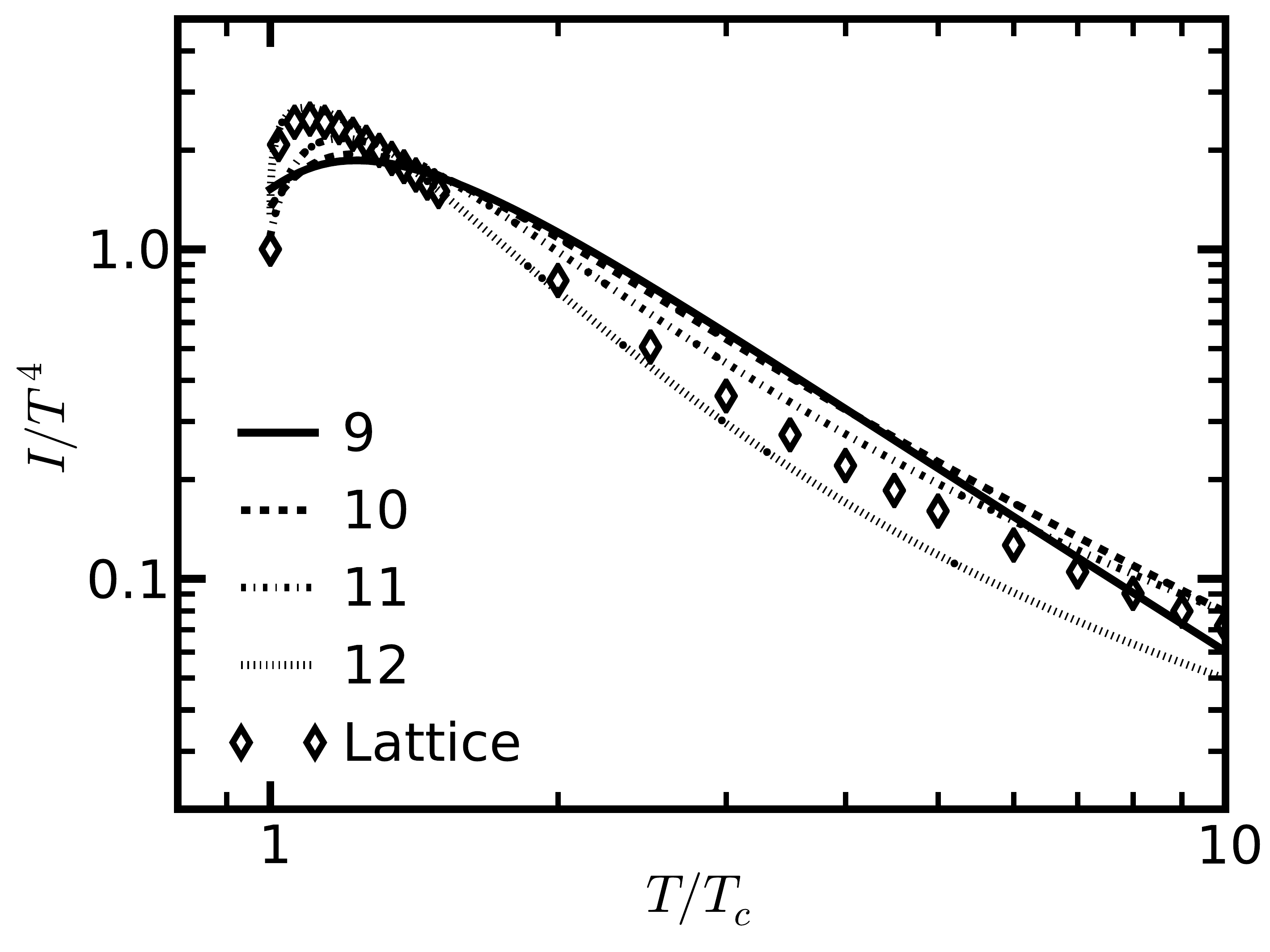}}
\includegraphics[width=0.327\columnwidth]{{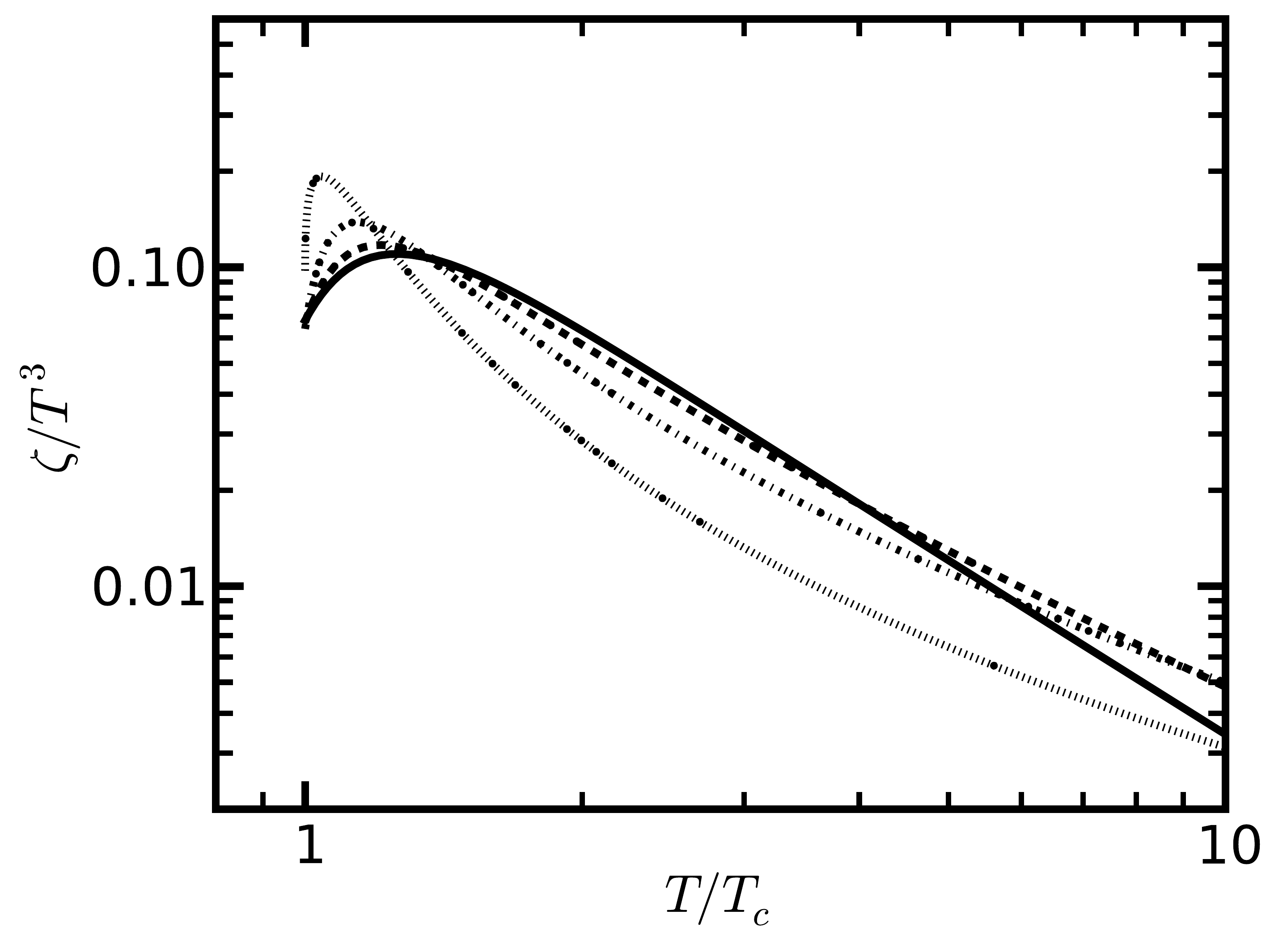}}
\includegraphics[width=0.327\columnwidth]{{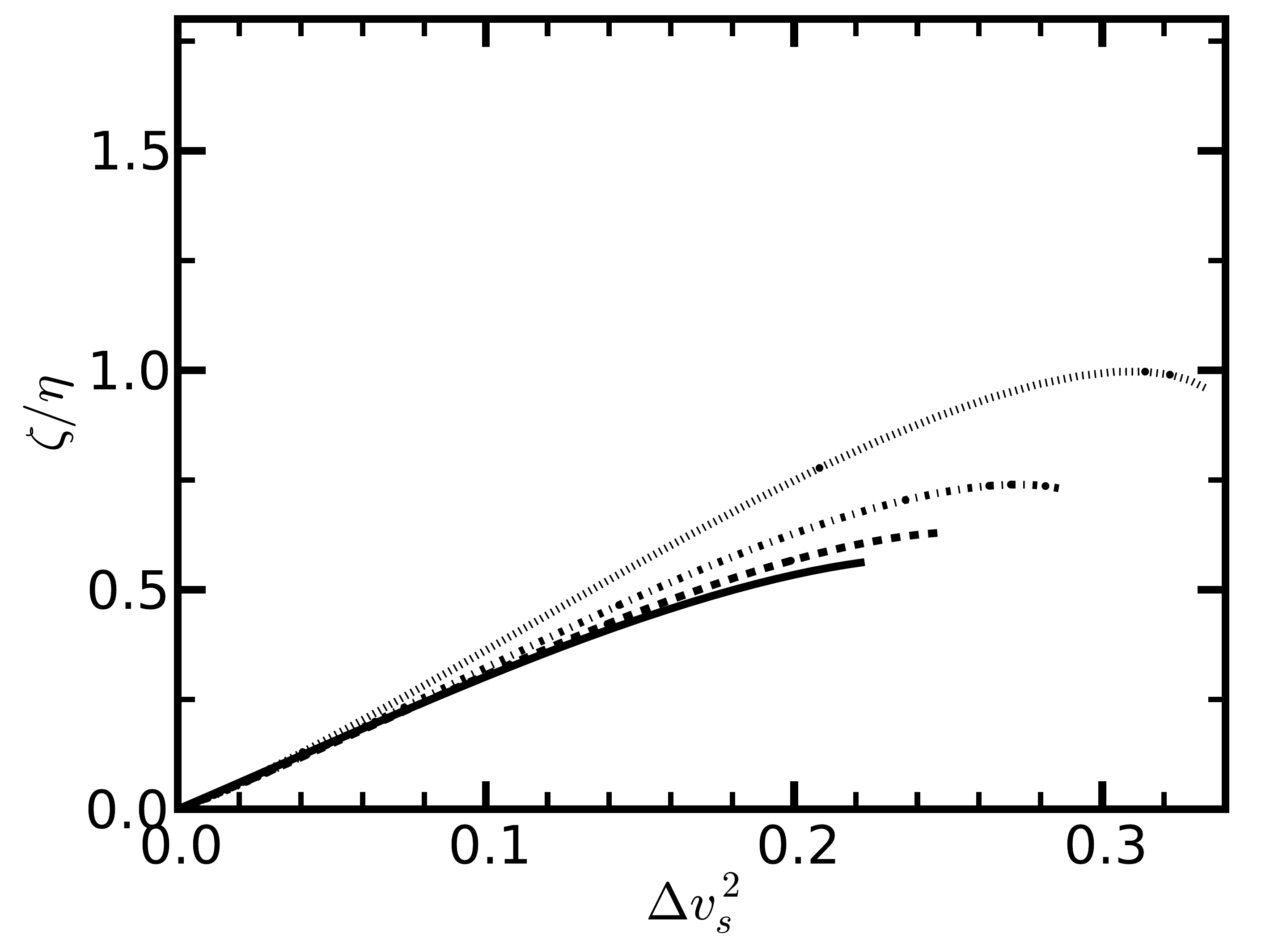}}
\end{center}

%
%

\caption{Equation of state $I/T^4$ as a function of temperature (left column),
scaled bulk viscosity $\zeta /T^3$ as a function of temperature (middle column)
and 
bulk to shear viscosity ratio as a function of non-conformality  
measure (right column). The numbers in the left panels refer to the
loci in the $\gamma$ vs.\ $\Delta $ plane in Fig.~\ref{fig:2}.    
}
\label{fig:3}
\end{figure} 

\section{Summary}

Despite of a lacking gravity dual to thermal SU(3) gauge theory, a gravity-scalar  
model with an appropriate ansatz for the potential allows for perfectly matching
of thermodynamics in the temperature region $(1 - 10) T_c$. Note that no
additional constraints are required, e.g.\ on scale settings or on the confined 
low-temperature phase or on the asymptotic behavior. The matching condition
forces the bulk to shear viscosity ratio to $\zeta / \eta = C \pi  \Delta v_s^2$ with
$C \approx 1.2$ for $\Delta v_s^2 < 0.25$, in agreement with a previously employed quasi-particle
model \cite{Bluhm} and the IHQCD model \cite{Kiritsis}. 
Without matching, the considered class of potentials exhibits 
significant variations of both $s/T^3$ and $\zeta/\eta$;
deviations from the linear relation $\zeta / \eta \propto \Delta v_s^2$ may occur 
over a larger range of $\Delta v_s^2$. The increase of $\zeta /\eta$ as a function
of the temperature toward $T_c$, however, seems to be a generic feature.
It is always less pronounced than the behavior found in \cite{Karsch}.
  
Our considerations ignore potentially strong curvature effects beyond the
classical gravity scenario, the reference to large 't Hooft coupling as well as a direct
link to the QCD $\beta$ function. In so far, we present an exploratory study
of a restricted set of observables in a special bottom-up set-up
leaving a systematic relation to the {\it ad hoc} employed AdS/CFT correspondence
with controlled deformation to accommodate the non-conformality for
further studies.   

The work is supported by BMBF grant 05P12CRGH1 and European Network HP3-PR1-TURHIC.

\end{document}